\documentclass[11pt]{article}
\usepackage{epsfig}
\usepackage{amsfonts}
\usepackage{bbm}
\newcommand{\la}[1]{\label{#1}}
\newcommand{\be}{\begin{equation}}
\newcommand{\ee}{\end{equation}}
\newcommand{\ba}{\begin{eqnarray}}
\newcommand{\ea}{\end{eqnarray}}
\newcommand{\rmi}[1]{{\mbox{\scriptsize #1}}}
\newcommand{\fig}{Fig.~}
\newcommand{\eq}{Eq.~}
\newcommand{\se}{Sec.~}
\newcommand{\eqs}{Eqs.~}
\newcommand{\nr}[1]{(\ref{#1})}
\newcommand{\tr}{{\rm Tr\,}}
\newcommand{\nn}{\nonumber \\}

\renewcommand{\vec}[1]{{\bf #1}}
\newcommand{\msbar}{{\overline{\mbox{\rm MS}}}}

\newcommand{\nG}{n_{\rm G}}

\newcommand{\bmu}{\bar\mu}
\def\lsi{\raise0.3ex\hbox{$<$\kern-0.75em\raise-1.1ex\hbox{$\sim$}}}
\def\gsi{\raise0.3ex\hbox{$>$\kern-0.75em\raise-1.1ex\hbox{$\sim$}}}

\newcommand{\unit}{{\mathbbm{1}}} 
\newcommand{\vmin}{v_\rmi{min}}

\makeatletter \@addtoreset{equation}{section} \makeatother

\makeatletter
\renewcommand\section{\@startsection {section}{1}{\z@}%
                                   {-5.5ex \@plus -1ex \@minus -.2ex}
                                   {2.3ex \@plus.2ex}%
                                   {\normalfont\large\bfseries}}
\renewcommand\subsection{\@startsection{subsection}{2}{\z@}%
                                     {-3.25ex\@plus -1ex \@minus -.2ex}%
                                     {1.5ex \@plus .2ex}%
                                     {\normalfont\normalsize\bfseries}}
\renewcommand\thesection {\@arabic\c@section}
\renewcommand\thesubsection   {\thesection.\@arabic\c@subsection}
\renewcommand{\@seccntformat}[1]{%
\csname the#1\endcsname.\hspace{1.0em}}
\makeatother

\begin{document}

\begin{titlepage}
\begin{flushright}
BI-TP 2005/48\\
hep-ph/0511246\\
\end{flushright}
\begin{centering}

\vfill

{\Large{\bf Baryon and lepton number violation rates \vspace*{1.5mm} \\
across the electroweak crossover}} 

\vspace{0.8cm}

Y.~Burnier$^\rmi{a}$,
M.~Laine$^\rmi{b}$, 
M.~Shaposhnikov$^\rmi{a}$

\vspace{0.8cm}

$^\rmi{a}${\em
Institut de Th\'eorie des Ph\'enom\`enes Physiques, EPFL, \\ 
CH-1015 Lausanne, Switzerland\\}

\vspace{0.3cm}

$^\rmi{b}${\em
Faculty of Physics, University of Bielefeld, 
D-33501 Bielefeld, Germany\\}

\vspace*{0.8cm}
 
\mbox{\bf Abstract}

\end{centering}

\vspace*{0.3cm}
 
\noindent
We point out that the results of many baryogenesis scenarios operating
at or below the TeV scale are rather sensitive to the rate of anomalous
fermion number violation across the electroweak crossover. 
Assuming the validity of the Standard Model of electroweak interactions, 
and making use of previous theoretical work at small Higgs masses, 
we estimate this rate for experimentally allowed values of the  
Higgs mass ($m_H = 100 ... 300$~GeV). We also elaborate on how the rate 
makes its appearance in (leptogenesis based) baryogenesis computations. 

\vfill


\vspace*{1cm}
 
\noindent
January 2006

\vfill

\end{titlepage}

%
\section{Introduction}

The scenario of thermal leptogenesis~\cite{fy} relies on anomalous
baryon + lepton number violation~\cite{gth},  which is very rapid at
temperatures above the electroweak scale~\cite{krs}, to convert the
original lepton asymmetry into an observable  baryon asymmetry.
Usually the temperature range where the lepton asymmetry generating
source terms are active, is much above the electroweak  scale. In
this case the anomalous processes have ample time to operate,  and
their precise rate is not important. In fact, the conversion factors
are simple analytic functions~\cite{ks,ls}, for which various
limiting values were derived already long ago~\cite{xs,all}. 

However, baryon asymmetry generation 
may also be a low temperature phenomenon,  
in which CP-breaking source terms are active down to
the electroweak scale;  for recent examples, see
Refs.~\cite{ars}--\cite{ch}. In this case the temperature dependence
of the anomalous rate does play an important role. This is even more
so for the large  (Standard Model like) Higgs masses that are
currently allowed  by experiment~\cite{pdg}: the electroweak symmetry
gets ``broken'' through an analytic crossover rather than  a sharp
phase transition~\cite{isthere,su2u1}, whereby the anomalous rate
also decreases only gradually.   

To allow for a precise study of generic scenarios of this type,
it is the purpose of this note to collect together all the 
relevant rate equations, such that systematic errors from 
this part of the computation can be brought under reasonable control.
We reiterate the baryon and lepton violation rate equations
in \se\ref{se:baryon}, 
estimate the anomalous ``sphaleron''
rate as a function of the Higgs mass and temperature
in \se\ref{se:sphaleron},
and summarise in \se\ref{se:summary}.

%
\section{Baryon and lepton number violation rates}
\la{se:baryon}

To zeroth order in neutrino Yukawa couplings, 
the Standard Model 
allows to define three global conserved charges:
\be
 X_i \equiv \frac{B}{\nG}  - L_i
 \;, 
\ee
where $B$ is the baryon number, $L_i$ the lepton number
of the $i^\rmi{th}$ generation, and $\nG$ denotes the 
number of generations. Given some values of $X_i$, a system
in full thermodynamic 
equilibrium at a temperature $T$ and with a Higgs 
expectation value $\vmin$ 
(suitably renormalised and in, say, the Landau gauge), 
contains then the baryon and lepton numbers~\cite{ks}
\ba
 && B \equiv B_\rmi{eq} \equiv 
 \chi\Bigl(\frac{\vmin}{T}\Bigr) \sum_{i=1}^{\nG} X_i
 \;, \quad
 L_i \equiv L_{i,\rmi{eq}} \equiv \frac{B_\rmi{eq}}{\nG} - X_i 
 \;, \la{Beq} \\
 && \chi(x) =   \frac{4[5 + 12 \nG + 4 \nG^2 + (9 + 6 \nG) x^2]}
                  {65 + 136 \nG + 44 \nG^2 + (117 + 72 \nG) x^2}
 \;. \la{chi}
\ea 
These relations hold up to corrections of order  
$\mathcal{O}((X_i/VT^3)^2)$ from the expansion in small chemical potentials,
$\mathcal{O}((h\vmin/\pi T)^2)$ from the high-temperature expansion, 
as well as $\mathcal{O}(h^2)$ from the weak-coupling expansion, 
where $h$ is a generic coupling constant. 

If we deviate slightly from thermodynamic equilibrium, 
the baryon and lepton numbers evolve with time. 
A non-trivial derivation~\cite{xs} yields the equations~\cite{xs,ks,rs}
\ba
 && \dot B(t) = -\nG^2 \,\rho\Bigl(\frac{\vmin}{T}\Bigr) 
 \frac{\Gamma_\rmi{diff}(T)}{T^3} [B(t) - B_\rmi{eq}] 
 \;, \quad \dot L_i(t) = \frac{\dot B(t)}{\nG} 
 \;, \la{Bdot} \\
 && 
 \rho(x) = \frac{3[65 + 136 \nG + 44 \nG^2 + (117 + 72 \nG)x^2]}
                {2\nG[30 + 62 \nG + 20 \nG^2 + (54 + 33\nG) x^2]}
 \;.
 \la{rho}
\ea
In the literature the factor 
$\nG^2\, \rho(\vmin/T)$ is often replaced with the constant
$13\nG/4$, which indeed is numerically an excellent approximation.
The term $\Gamma_\rmi{diff}(T)$ is called the 
Chern-Simons diffusion rate, 
or (twice) the sphaleron rate, and is defined by
\be
 \Gamma_\rmi{diff}(T) \equiv \lim_{V,t\to\infty}
 \frac{\langle Q^2(t) \rangle_T}{V t}
 \la{Gammadef}
 \;,
\ee
where $Q(t) \equiv \int_0^t \! {\rm d}t' \int_V \! {\rm d}^3 \vec{x}' \, q(x') 
\equiv N_{\mbox{\tiny\rm{CS}}}(t) - N_{\mbox{\tiny\rm{CS}}}(0)$ 
is the topological charge, 
and $N_{\mbox{\tiny\rm{CS}}}(t)$ is the Chern-Simons number. The 
expectation value in \eq\nr{Gammadef} is to be evaluated in 
a theory without fermions~\cite{xs}. Corrections to \eq\nr{rho}
are of the same type as those to \eq\nr{chi}.

For practical purposes, it is useful to eliminate the conserved
charges $X_i$ from the equations, and write just a coupled system 
for $B(t), L_i(t)$. Defining 
\be
 \gamma \equiv \nG^2 \, \rho\Bigl(\frac{\vmin}{T}\Bigr) 
 \Bigl[ 1 - \chi\Bigl( \frac{\vmin}{T} \Bigr) \Bigr] 
 \frac{\Gamma_\rmi{diff}(T)}{T^3} \;, \quad
 \eta \equiv \frac{\chi({\vmin}/{T})}{1-\chi({\vmin}/{T})} \;, \la{gammadef}
\ee 
and introducing sources $f_i(t)$ for the lepton numbers, 
we can convert \eqs\nr{Beq}, \nr{Bdot} to
\ba
 \dot B(t) & = & - \gamma(t) 
 \Bigl[ B(t) + \eta(t) \sum_{i=1}^{\nG} L_i(t) \Bigr] \;, \la{dot1} \\
 \dot L_i(t) & = & -\frac{\gamma(t)}{\nG} 
 \Bigl[B(t) + \eta(t) \sum_{i=1}^{\nG} L_i(t)\Bigr] 
 + f_i(t)
 \;. \la{dot2}
\ea

These equations can easily be integrated, if we know the temperature
dependence of $\vmin/T$ and the time evolution of $T$. The solution is 
particularly simple if we make use of the fact that
$\eta$ is, to a reasonable approximation, a constant, 
$\eta(t) \simeq 0.52 \pm 0.03$. In this case linear combinations
of \eqs\nr{dot1}, \nr{dot2} yield independent first order equations
for $B(t)-L(t)$ and $B(t) + \eta L(t)$, 
where $L(t) \equiv \sum_{i=1}^{\nG} L_i(t)$. Denoting
$\omega(t';t) \equiv \exp[-(1+\eta) \int_{t'}^t \! {\rm d} t'' \,  
\gamma(t'')]$ and $f(t) \equiv \sum_{i=1}^{\nG} f_i(t)$, the solution reads
\be
 B(t) = \frac{1}{1+\eta} \biggl\{ 
 \Bigl[ B(t_0) + \eta L(t_0) \Bigr] \omega(t_0;t) + \eta
 \Bigl[ B(t_0) - L(t_0) \Bigr] 
 - \eta \int_{t_0}^t \! {\rm d} t' \, 
 f(t') \Bigl[ 1 - \omega(t';t) \Bigr] 
 \biggr\} 
 \;. 
 \la{soln}
\ee
A further simplification follows by noting that 
$\omega(t';t)$ varies very rapidly with the time $t'$ 
around a certain $t' \sim t_*$, from zero at $t' < t_*$ to 
unity at $t' > t_*$, 
while $f(t')$ is a slowly varying function of time. Assuming
furthermore that $B(t_0) = L(t_0) = 0$, we obtain
\be
 B(t) \approx \frac{-\eta}{1+\eta} 
 \int_{t_0}^{t_*} \! {\rm d} t' \, f(t') = 
 - \chi \int_{t_0}^{t_*} \! {\rm d} t' \, f(t')
 \;, \la{Bappro}
\ee
where the ``decoupling time'' can be defined as 
$t_* \equiv t_0 + \int_{t_0}^t \! {\rm d} t' \, 
[1-\omega(t';t)]$.
Thus, if $f(t')\neq 0$ around the time $t_*$, the baryon 
asymmetry generated depends sensitively on $t_*$, and it is 
important to know the function 
$\omega(t';t)$, 
determined by $\gamma(t'')$, quite precisely. 

The equations that we have written were formally derived in Minkowski 
spacetime. They are easily generalised to an expanding background, 
however: their form remains invariant if we simply replace the total
(comoving) baryon and lepton numbers
$B$, $L_i$ by number densities over the entropy density $s(T)$:
$B \to n_B \equiv B/[a^3 s(T)]$,
$L \to n_L \equiv L/[a^3 s(T)]$,
where $a^3$ is a comoving volume element.  
Furthermore, it is often convenient to replace
time derivatives with temperature derivatives via 
\be
 \frac{{\rm d}}{{\rm d}t} = -\frac{\sqrt{24\pi}}{m_\rmi{Pl}}
 \frac{\sqrt{e(T)}}{{\rm d}[\ln s(T)]/{\rm d}T} \frac{{\rm d}}{{\rm d}T}
 \;,
\ee
where $e(T)$ is the energy density; 
we assumed the Universe to be flat ($k=0$); 
and we ignored the cosmological constant. 
Both $s(T) = p'(T)$  and $e(T)= T s(T) - p(T)$
follow from the thermodynamic pressure $p(T)$
which is known to high accuracy~\cite{gv}, but can 
in practice be reasonably well approximated with the ideal gas formula
$p(T) \approx g_* \pi^2 T^4/90$, with $g_*\simeq 106.75$.

In many baryogenesis scenarios, the source terms $f_i(t)$ 
in \eq\nr{dot2} are 
approximated by Boltzmann-type equations for the various left-handed
and right-handed neutrino number densities. Collecting the number
densities to the matrices $\vec{n}_L, \vec{n}_R$, respectively, with the 
normalization $\tr[\vec{n}_L] = n_L$, a concrete realization of 
\eqs\nr{dot1}, \nr{dot2} could then read 
\ba
 \dot n_B(t)   & = & -\gamma(t) 
 \Bigl\{ n_B(t) + \eta(t) \tr[\vec{n}_L(t)] \Bigr\} 
 \;, \la{ndot1} \\
 \dot \vec{n}_L(t) & = & -\frac{\gamma(t)}{\nG} \Bigl\{ n_B(t) + \eta(t) 
  \tr[\vec{n}_L(t)] \Bigr\} \raise-0.2mm\hbox{$\unit$} + 
                 \mathcal{F}_L[\vec{n}_R,\vec{n}_L,t] 
 \;, \\[1.5mm]
 \dot \vec{n}_R(t) & = & \mathcal{F}_R[\vec{n}_R,\vec{n}_L,t] 
 \;, \la{ndot3}
\ea 
with functionals $\mathcal{F}_L, \mathcal{F}_R$ that need to 
be determined for the specific model in question.

%
\section{Chern-Simons diffusion rate}
\la{se:sphaleron}

An essential role in the rate equations \nr{ndot1}--\nr{ndot3}
is played by the function $\gamma(t)$ whose time dependence is, 
via \eq\nr{gammadef}, dominantly determined by $\Gamma_\rmi{diff}(T)$,
defined in \eq\nr{Gammadef}. We now collect together
the current knowledge concerning $\Gamma_\rmi{diff}(T)$
in the Standard Model.

At high temperatures (in the ``symmetric phase'') the Chern-Simons
diffusion rate is purely non-perturbative, and needs to be evaluated
numerically. So-called classical real-time simulations~\cite{a} 
produce $\Gamma_\rmi{diff}(T) = (25.4\pm 2.0) \alpha_w^5 T^4$~\cite{bmr}, 
where the number 25.4 is in fact the value of a function containing 
terms like $\ln(1/\alpha_w)$~\cite{db}, at the physical $\alpha_w$.

At lower temperatures, the rate is traditionally written 
in the form~\cite{aml}
\be
 \Gamma_\rmi{diff}(T) = 
 4 T^4 \frac{\omega_-}{gv_\rmi{min}} 
 \biggl( \frac{\alpha_w}{4\pi} \biggr)^4
 \biggl( \frac{4\pi v_\rmi{min}}{g T} \biggr)^7
 \mathcal{N}_\rmi{tr} 
 (\mathcal{N}\mathcal{V})_\rmi{rot}
 \; \kappa \; \exp\biggl( - \frac{E_\rmi{sph}}{T}\biggr)
 \;. \la{Gammadiff}
\ee
Here $g$ is the SU(2) gauge coupling, 
$\alpha_w = g^2/4\pi$; 
$\omega_-$ might generically be called the dynamical prefactor, 
and is related to the absolute value of 
the negative eigenvalue of the fluctuation operator 
around the sphaleron solution; 
$\mathcal{N}_\rmi{tr} 
 (\mathcal{N}\mathcal{V})_\rmi{rot}$
are normalisation factors related to the zero-modes of the 
fluctuation operator; 
$\kappa$ contains the contributions of the positive modes; 
and $E_\rmi{sph}$ is the energy of the saddle-point configuration
(the sphaleron)~\cite{km}.

Most of the factors appearing in \eq\nr{Gammadiff} have been 
evaluated long ago. In particular, $E_\rmi{sph}$ can be found
in Ref.~\cite{sinW} for the bosonic sector of the SU(2)$\times$U(1)
Standard Model, while
fermionic effects were clarified in Ref.~\cite{gdm}. 
The zero-mode factors and (the naive version of) $\omega_-$
were evaluated in Refs.~\cite{ahy,cml}, while  $\kappa$
was determined numerically in Refs.~\cite{cmlw,bj}.

Unfortunately, it as not {\it a priori} clear how accurate
the corresponding results are. Indeed, \eq\nr{Gammadiff} has 
an inherently 1-loop structure, but it is known from studies
of the electroweak phase transition that 2-loop effects, 
parametrically suppressed only by the infrared-sensitive 
expansion parameter $\mathcal{O}(hT/\pi \vmin)$, are  
large in practice~\cite{ae,pert}. Moreover,
the naive definition of $\omega_-$ through the 
negative eigenvalue does not appear to be correct~\cite{asy}.

A reliable determination of $\Gamma_\rmi{diff}$
can again be obtained by numerical methods, employing real-time classical 
simulations. Of course classical simulations are not
exact either, but they do contain the correct infrared physics, 
and should thus only suffer from infrared-safe errors   
of the type mentioned below \eq\nr{chi}. Thus, classical
simulations allow in principle to incorporate 
the dominant higher order effects,
as well as a correct treatment of $\omega_-$. 

In the ``broken symmetry phase'', 
large-scale classical simulations have been carried out 
in Ref.~\cite{gdm_b}.
Unfortunately, they
only extend up to Higgs masses around $m_H = 50$~GeV, 
and were only carried out for certain temperatures
(there are some results also at larger Higgs masses but
with less systematics~\cite{sd}). While
we have not carried out any new simulations, we do make use 
of the observation~\cite{gdm_b} that the discrepancy
between the numerical results, and a certain analytical recipe, 
of the type reiterated below, appears to be independent 
of the Higgs mass. We thus extend the analytical
recipe to large Higgs masses, 
and add to these results a (small) constant correction factor, 
extracted from Ref.~\cite{gdm_b}. 
In practice, the steps are as follows:
 
\vspace*{0.3cm}

(i)
We employ the (resummed) 2-loop finite-temperature 
effective potential $V(v)$ in Landau 
gauge, as it is specified in Ref.~\cite{generic}. Effects of 
the hypercharge group U(1) need to be taken into account
only at 1-loop level, as demonstrated in Ref.~\cite{su2u1}. The potential
is parametrised by the zero-temperature physical quantities 
$m_W$, $m_Z$, $m_\rmi{top}$, $m_H$, $\alpha_s(m_Z)$, $G_F$; 
their values (apart from $m_H$) are taken from Ref.~\cite{pdg}.

We remark that
although this is formally a higher order effect, the effective potential 
does depend on the scale parameter $\bmu$ of the $\msbar$ scheme. 
One may thus consider various
choices of $\bmu$. We follow a strategy 
similar to Ref.~\cite{pert} and write
$
 V(v) - V(0) = \int_0^v \! 
 {\rm d} v' \left. 
 \partial V(v') / 
 \partial v'  \right|_{\bmu = \bmu(v')}
$,
where the scale is chosen as 
$\bmu(v) \equiv \Delta\, \sqrt{3 \lambda_\rmi{eff} v^2}$,
where $\lambda_\rmi{eff}$ is the scalar coupling
of the dimensionally reduced theory~\cite{generic} and 
$\Delta$ is a constant. We consider 
$\Delta \equiv 1.0$ as the 
``reference value'', while variations in the range 
$\Delta = 0.25 ... 4.0$ indicate the magnitude of uncertainties. 
 
\vspace*{0.3cm}

(ii)
To avoid threshold singularities at small $v$ related to the Higgs 
and Goldstone masses, we replace the exact 2-loop potential 
by a polynomial fit around the broken minimum: 
\be
  \frac{\mathop{\mbox{Re}} [V(v) - V(0)]}{T^4} = 
 \sum_{n = 2}^4 b_n \, (\hat v - \hat v_\rmi{min})^n + 
 \mathcal{O}((\hat v - \hat v_\rmi{min})^5)
 \;,  
\ee
where $\hat v \equiv v/T$. We carry out the fit 
in the range $v = (0 ... 1.5)\, v_\rmi{min}$.
Only values $v \le v_\rmi{min}$ are needed 
for the sphaleron solution, but including some larger values
allows for a better fit of the curvature around the minimum.
We have considered other fit forms as well and find that the errors
introduced through the fitting are insignificant compared
with other error sources.

\vspace*{0.3cm}

(iii)
We compute the sphaleron energy $E_\rmi{sph}/T$
for this potential. We assume that the use of the 2-loop potential
rather than the tree-level potential takes care of the factor $\kappa$
in \eq\nr{Gammadiff}, which we thus set to unity. At 1-loop level
this can to some extent be demonstrated explicitly~\cite{bj}, but 
what is more important for us is that any possible errors from
this approximation are compensated for by step (v) below.
The effect of the U(1)  group
is treated perturbatively~\cite{km}, which is an excellent 
approximation~\cite{sinW}. We use an effective finite-temperature
Weinberg-angle $\tan^2(\theta_W)_\rmi{eff} \approx 0.315$~\cite{su2u1}.

\vspace*{0.3cm}

(iv)
We determine the zero-mode factors
$\mathcal{N}_\rmi{tr}$, $(\mathcal{NV})_\rmi{rot}$ and
the dynamical factor $\omega_-$, 
as described in Ref.~\cite{cml}, except that every 
appearance of the tree-level $\lambda(h^2-1)^2/4g^2$ is 
replaced by the 2-loop potential 
$V(h v_\rmi{min})/g_\rmi{eff}^2v_\rmi{min}^4$.
We also determine the effective gauge coupling
$g_\rmi{eff}$ of the dimensionally reduced theory~\cite{generic}, 
and use $g_\rmi{eff}$ instead of $g$ in \eq\nr{Gammadiff}.
The effect of the zero-mode factors and $\omega_-$ is to 
effectively decrease $E_\rmi{sph}/T$ by about 15\%, 
or by $3...10$ in absolute units. 

\vspace*{0.3cm}

(v)
Finally we add a correction from Ref.~\cite{gdm_b}, 
which we assume to be a constant:
\be
 \Gamma_\rmi{diff}^\rmi{(full)} \equiv \Gamma_\rmi{diff}^\rmi{(i)-(iv)} 
 \exp\Bigl[-(3.6 \pm 0.6)\Bigr]
 \;. \la{gdm}
\ee
This correction is in most cases subleading compared 
with those in step (iv), and goes in the opposite direction. 
It may be noted that 
there is some latitude with respect to which gauge 
is used for the evaluation of the prefactors appearing
in \eq\nr{Gammadiff}~\cite{aml,cml}, but since
$\Gamma_\rmi{diff}^\rmi{(full)}$ is gauge-independent, the non-perturbative
correction factor compensates for this as well.

\vspace*{0.3cm}

(vi) 
Finally, since we rely on an extrapolation of 
the non-perturbative correction factor to larger Higgs masses, 
we assign a generous overall uncertainty to $\Gamma_\rmi{diff}$, 
in the range 
\be
   \left| \delta \ln \biggl[ \frac{\Gamma_\rmi{diff}(T)}{T^4} \biggr] 
   \right | 
   \approx 2.0 
   \;. 
  \la{Gammaerror}
\ee
This amounts to roughly three times the error in \eq\nr{gdm}.
We stress that even though the Higgs masses leading to \eq\nr{gdm} 
are much smaller than we consider, 
the values of $v_\rmi{min}/T$ are similar,
and thus the bulk of the effect in \eq\nr{gdm} should still 
remain intact, at least in the physically most plausible range 
100~GeV $\le m_H \le$ 200~GeV.

\vspace*{0.3cm}

\begin{figure}[t]


\centerline{%
\epsfysize=8.0cm\epsfbox{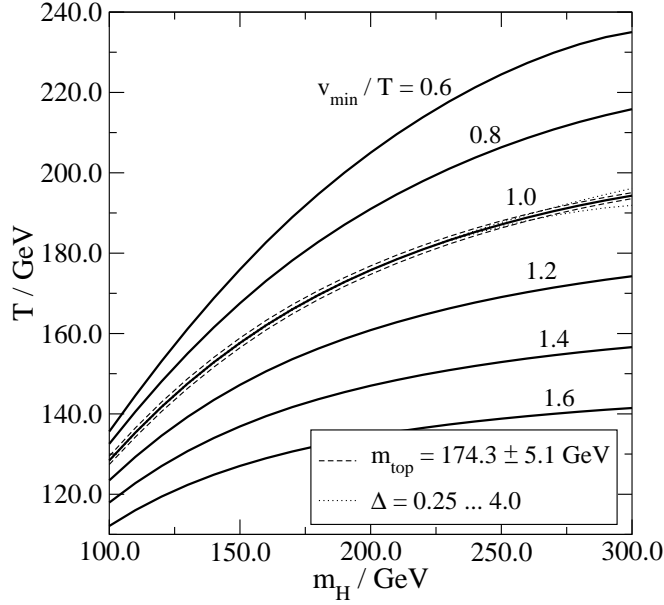}%
}

\vspace*{0.5cm}


\caption[a]{\it
The temperatures for which specific values of $v_\rmi{min}/T$
(in Landau gauge) are reached, as a function of the Higgs mass $m_H$. 
For $v_\rmi{min}/T = 1.0$ we also show the effects of the variations
$m_\rmi{top} = 174.3 \pm 5.1$~GeV (dashed lines) and
$\Delta = 0.25 ... 4.0$ (dotted lines).}

\la{fig:vev}

\end{figure}

In \fig\ref{fig:vev}, we show the location of the minimum 
of the 2-loop effective potential. We only consider values
for which the infrared sensitive expansion parameter 
$h T/\pi v_\rmi{min}$ remains reasonably small. For higher
temperatures, the corresponding rate $\Gamma_\rmi{diff}$ 
extrapolates smoothly to the symmetric phase value~\cite{sd}, 
like standard thermodynamic observables~\cite{isthere,su2u1,ag}.

\begin{figure}[t]


\centerline{%
\epsfysize=8.0cm\epsfbox{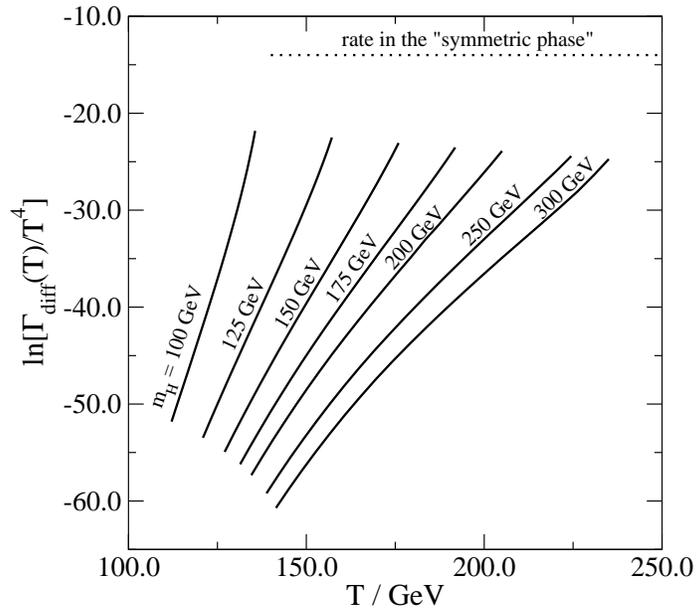}%
}

\vspace*{0.5cm}


\caption[a]{\it
$\ln[\Gamma_\rmi{diff}(T)/T^4]$ as 
a function of the Higgs mass and temperature.
The overall error is estimated in \eq\nr{Gammaerror}. 
The dotted horizontal line indicates the value which 
all curves approach at large $T$.
The values in the range 100~GeV $\le m_H \le$ 200~GeV
can be roughly approximated by \eq\nr{fit}.
Note that the rate falls off more slowly at large Higgs masses.}

\la{fig:rate}

\end{figure}

The rates $\Gamma_\rmi{diff}$ are displayed
in~\fig\ref{fig:rate}, with assumed uncertainties of the order
in \eq\nr{Gammaerror}. For practical applications, we note that
in the range 100~GeV $\le m_H \le$ 200~GeV and for $T$ such that
$-\ln[\Gamma_\rmi{diff}(T)/T^4] \approx 30...50$,
the results can within our uncertainties be
approximated by
\be
 - \ln \biggl[\frac{\Gamma_\rmi{diff}(T)}{T^4} \biggr] 
 \approx \sum_{i,j\ge 0}^{i+j \le 2} c_{ij} 
 \biggl( \frac{m_H - 150\;\mbox{GeV}}{10\;\mbox{GeV}} \biggr)^i 
 \biggl( \frac{T - 150\;\mbox{GeV}}{10\;\mbox{GeV}} \biggr)^j 
 \;, \la{fit}
\ee
with the coefficients
\ba
 &&  c_{00} = 39.6
 \;, \quad  
 c_{10} = 3.52
 \;, \quad  
 c_{01} = -7.09
 \;, \nn
 && c_{20} = -0.376
 \;, \quad  
 c_{11} = 0.421
 \;, \quad  
 c_{02} = 0.170
 \;. \la{cs}
\ea

\begin{figure}[t]


\centerline{%
\epsfysize=8.0cm\epsfbox{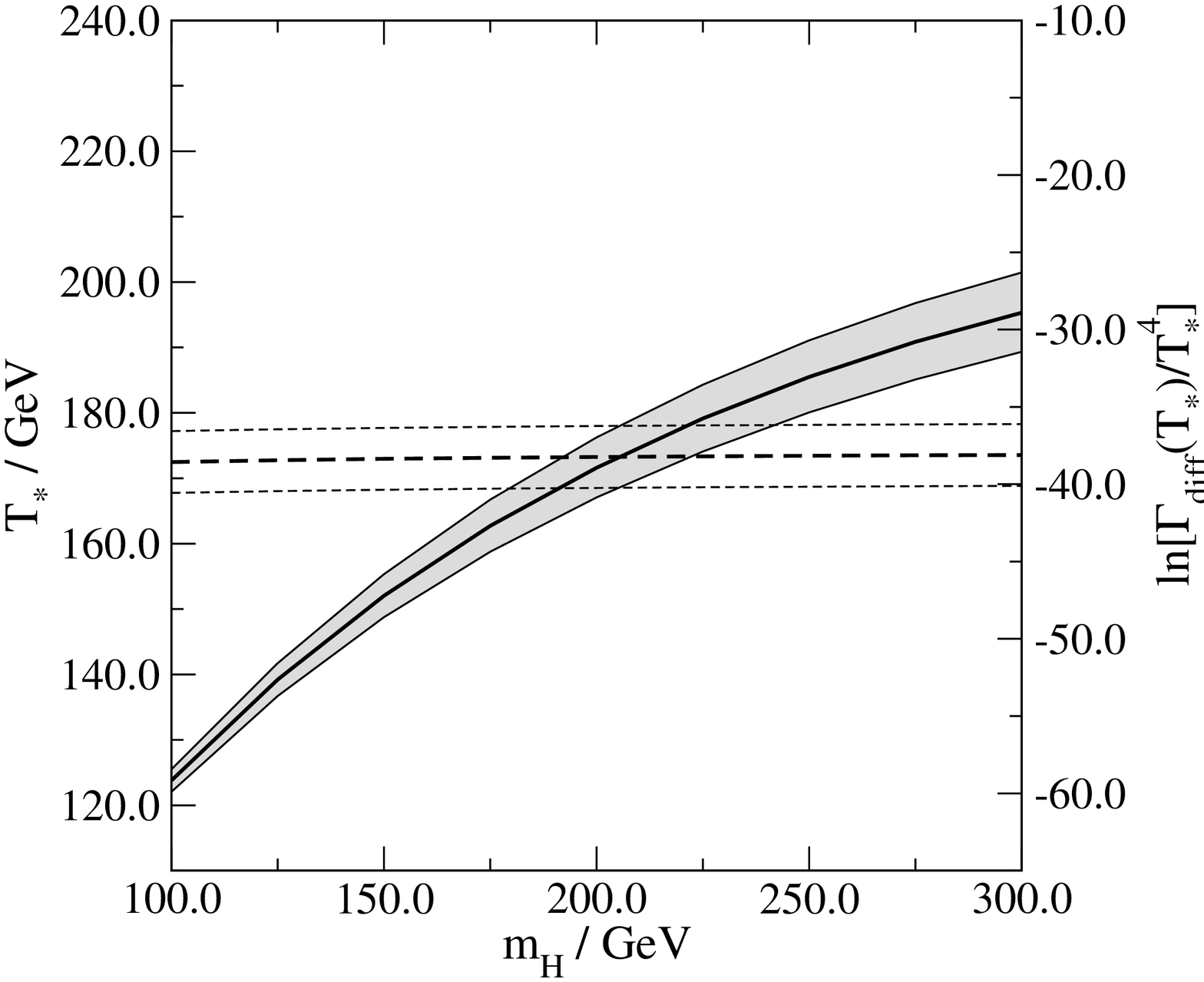}%
}

\vspace*{0.5cm}


\caption[a]{\it
The solid line indicates the decoupling temperature $T_*$ 
as defined in the text (assuming a constant $g_* \simeq 106.75$),
with an error band following from 
changing $\Gamma_\rmi{diff}(T_*)/T_*^4$ within the range
of \eq\nr{Gammaerror}. The dashed lines show
the corresponding anomalous rate.}

\la{fig:Tstar}

\end{figure}

Given $\Gamma_\rmi{diff}(T)/T^4$, we can finally estimate
the decoupling time $t_*$ and/or the corresponding decoupling
temperature $T_*$, needed in \eq\nr{Bappro}. In the limit that
$\Gamma_\rmi{diff}(T)/T^4$ changes very rapidly with $T$, 
the solution is given by the equation 
$\nG^2\,\rho\, \Gamma_\rmi{diff}(T_*)/T_*^3 = H(T_*)$, 
where $H(T)$  is the Hubble rate defined through
$H^2(T) = {8 \pi e(T)/ 3 m_\rmi{Pl}^2}$. 
Writing 
\be
 \ln \biggl[ \frac{\Gamma_\rmi{diff}(T)}{T^4} \biggr] = 
 \ln\biggl[ \frac{\Gamma_\rmi{diff}(T_*)}{T_*^4} \biggr]
 + A (T - T_*) + \mathcal{O}((T-T_*)^2)
 \;,
\ee
corrections to this leading order approximation are of relative
order $\mathcal{O}(1/ A T_*)$, which according to \eqs\nr{cs}
is in the one percent range, and thus subdominant compared
with other error sources. The leading order solution is shown 
in~\fig\ref{fig:Tstar}.

\begin{figure}[t]


\centerline{%
\epsfysize=8.0cm\epsfbox{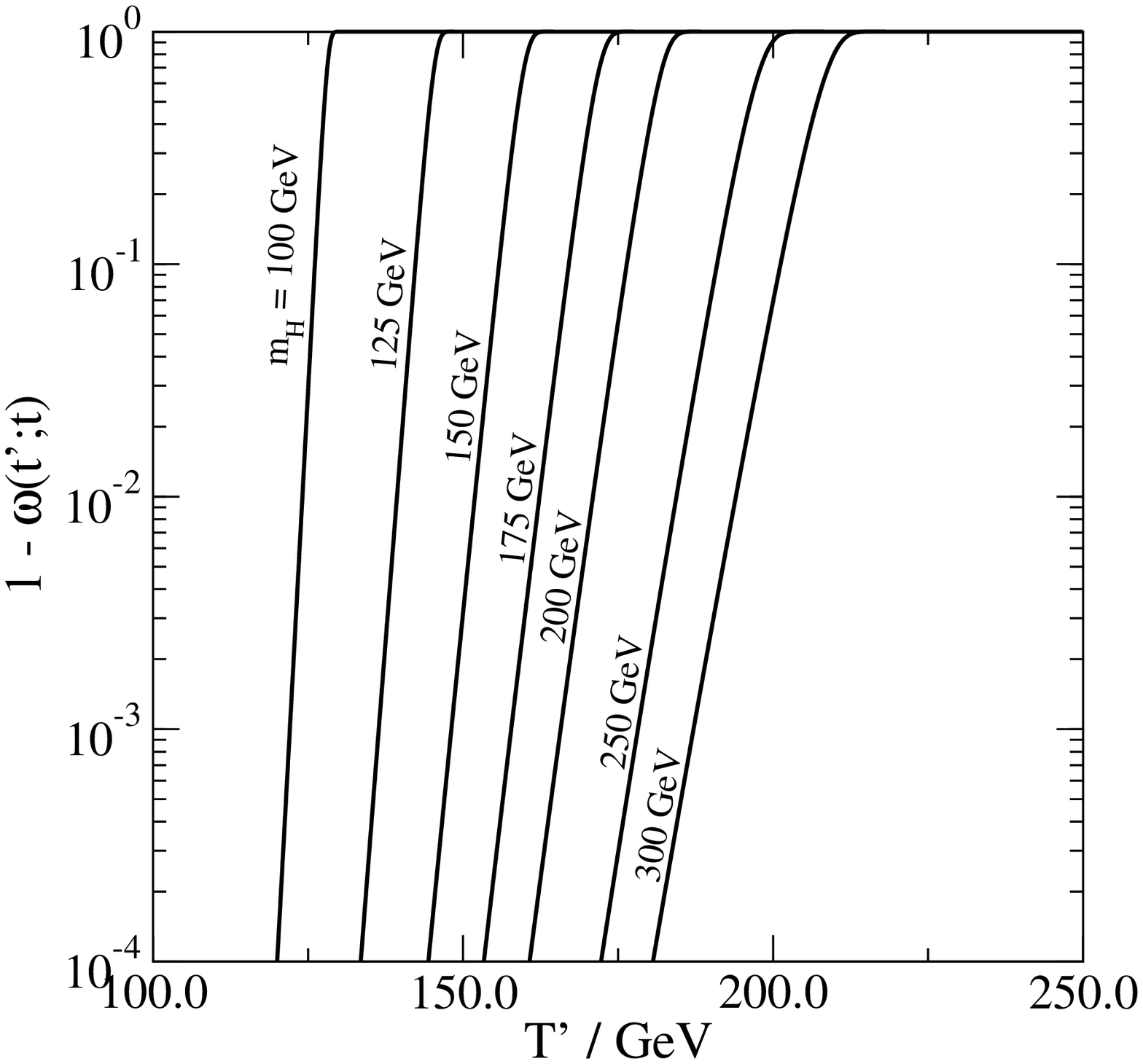}%
}

\vspace*{0.5cm}


\caption[a]{\it
The function $1-\omega(t';t)$ appearing in \eq\nr{soln}, 
as a function of the temperature $T'$ corresponding to the time $t'$ 
(the final moment $t$ is fixed to the point where $T=100$~GeV).
We indicate temperatures instead of times, because this significantly 
reduces the dependence on the constant $g_* \simeq 106.75$, which 
has non-negligible radiative corrections~\cite{gv}. This figure 
can be used to gauge the accuracy of the sudden decoupling
approximation shown in \fig\ref{fig:Tstar}.}

\la{fig:omega}

\end{figure}

Comparing \fig\ref{fig:Tstar} with \fig\ref{fig:vev}, it is seen
that $T_*$ corresponds to values $v_\rmi{min}/T = 1.0...1.2$. At the 
same time, the rate of change of $\Gamma_\rmi{diff}$ is less
abrupt ($A$ is smaller) at large Higgs masses, and a sudden 
decoupling is a less precise approximation. This can be seen
in \fig\ref{fig:omega}, where the full function 
$1-\omega(t';t)$ appearing in \eq\nr{soln}
is plotted. 

%
\section{Summary and conclusions}
\la{se:summary}

The main contents of this note are the baryon and lepton
number rate equations shown in \eqs\nr{gammadef}--\nr{dot2}, 
as well as the ``sphaleron rate'' $\Gamma_\rmi{diff}(T)/T^4$
that enters these equations, shown in \fig\ref{fig:rate}
and in \eq\nr{fit}. With this knowledge,
and given that the factors $\chi$, $\rho$, $\eta$ are to a fairly
good approximation constants, the equations can be 
integrated in closed form, leading to \eq\nr{soln}. 
An even simpler estimate for the baryon number generated 
in a given scenario 
can be obtained from \eq\nr{Bappro}, 
where $t_*$ corresponds to the temperature $T_*$ 
shown in \fig\ref{fig:Tstar}. On the other hand, 
the most precise results can be obtained by  
integrating \eqs\nr{gammadef}--\nr{dot2}  numerically
down to temperatures shown in \fig\ref{fig:rate}.
All of these equations are model-independent in form; 
the specific model enters through the source terms $f_i$.

The biggest uncertainties of our estimates for
$\Gamma_\rmi{diff}(T)/T^4$ originate from the fact that
systematic numerical studies have only been carried out
at fairly small Higgs masses~\cite{gdm_b,sd}.
If a Standard Model like Higgs particle is found
at the LHC, there is certainly a strong motivation 
for repeating the numerical studies at the physical value 
of the Higgs mass, in order to remove the corresponding
error source (\eq\nr{Gammaerror}) from our estimates.

%
\section*{Acknowledgements}
This work was partly supported by the Swiss Science Foundation.

\newpage


\end{document}